# Using an Expert Panel to Validate the Malaysian SMEs-Software Process Improvement Model (MSME-SPI)

Malek A. Almomani[1], Shuib Basri[2], Omar Almomani[3], Luiz Fernando Capretz[4], Abdullateef Balogun[2,5], Moath Husni[1] Abdul Rehman Gilal[6]

[1]Department of Software Engineering, The World Islamic Sciences and Education University, Amman, 11947, Jordan
{malek.almomani, moath.tarawneh}@wise.edu.jo

[2]Department of Computer and Information Sciences, Universiti Teknologi PETRONAS, Seri Iskandar, Perak, 32610 Malaysia
{shuib_basri, abdullateef_16005851}@utp.edu.my

[3]Department of Computer Network and Information Systems, The World Islamic Sciences and Education University, Amman, 11947, Jordan
omar.almomani@wise.edu.jo

[4]Department of Electrical and Computer Engineering, Western University, London, Ontario, Canada.
lcapretz@uwo.ca

[5]Department of Computer Science, University of Ilorin, PMB 1515, Ilorin, Nigeria.
balogun.ao1@unilorin.edu.ng

[6]Department of Computer Science, Sukkur IBA University, 65200, Pakistan
a-rehman@iba-suk.edu.pk

**Abstract.** This paper presents the components of a newly developed Malaysian SMEs - Software Process Improvement model (MSME-SPI) that can assess SMEs software development industry in managing and improving their software processes capability. The MSME-SPI is developed in response to practitioner needs that were highlighted in an empirical study with the Malaysian SME software development industry. After the model development, there is a need for independent feedback to show that the model meets its objectives. Consequently, the validation phase is performed by involving a group of software process improvement experts in examining the MSME-SPI model components. Besides, the effectiveness of the MSME-SPI model is validated using an expert panel. Three criteria were used to evaluate the effectiveness of the model namely: usefulness, verifiability, and structure. The results show the model effective to be used by SMEs with minor modifications. The validation phase contributes towards a better understanding and use of the MSME-SPI model by the practitioners in the field.

# 1 Introduction

Software Process Improvement (SPI) models are developed to improve the development process within companies and it mainly oriented towards large companies [17]. The SPI models are very difficult to be adopted in the Small and Medium software Enterprises (SMEs) industry because it requires large software process details [17,18]. In the Malaysian context, several studies indicated that Malaysian SME software development companies face myriad challenges to implement SPI efforts [17-20]. However, implementing SPI amongst SMEs software development industry still possible by harnessing their relative strength of the affective factors [18,21,22]. Therefore, Malaysian SMEs-Software Process Improvement (MSME-SPI) has been developed specifically for the software development industry. Thereby, the development of the model as explained in Fig. 1, illustrated the stages of MSME-SPI which were developed based on existing models such as (CMMI Dev 1.3, ISO/IEC 15504, MoProSoft, IMM, and SOVRM). Success Factors (SFs), Barriers Factors (BFs), and their corresponding best practices concepts are integrated with generic elements of SPI as presented in Figure 1. Accordingly, the MSME-SPI is required to be verified for several issues. Firstly, verify the usefulness of the model for software companies in assessing the strength and weaknesses of their points. Secondly, verifying the verifiability of the MSME-SPI practices to be useful for the staff implementing them and other members of the companies. Thirdly, verifying the suitability of MSME-SPI structure such that the distribution of critical factors (SFs and BFs) among different levels. Finally, verifying best practices reviewed that design under each identified critical factors to cover all related practices for any modification or further changes. However, case studies have been used to verify MSME-SPI [16].

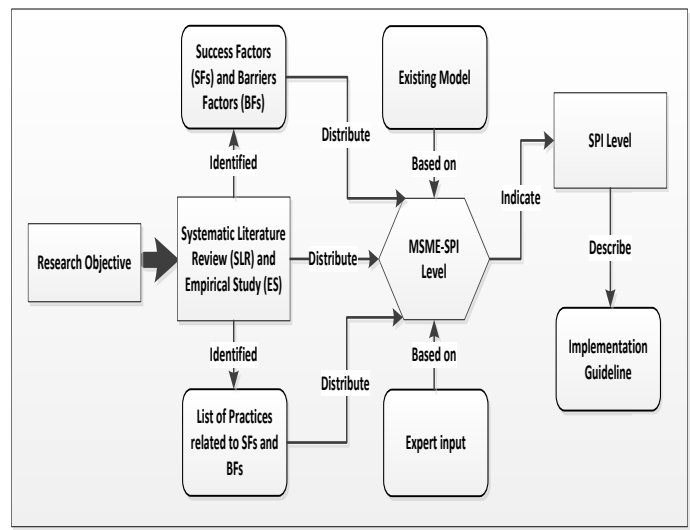

**Fig. 1.** Structure of the Malaysian SMEs - Software Process Improvement model.

The main objective of this phase is to validate and assess the effectiveness of the MSME-SPI model amongst the practitioners in the software development industry. Hence, the validation process is important to confirm the validity of the MSME-SPI model. It aims to show how the MSME-SPI model can support the Malaysian SME software development companies in managing and improving their software processes capabilities. Moreover, validation would facilitate enhancements and refinement of the MSME-SPI model.

This research paper has been organized as follows. Section 2 presents and explains the research methodology employed to achieve the research objectives. In Section 3, the results and findings of the research study are briefly reported. Finally, Section 4 presents the conclusions of the study, which includes directions and suggestions for future work.

## 2      Research Methodology

In this research, to validate the MSME-SPI model semi-structured interview-based validation approach was designed. The main reason for carrying out a semi-structured interview approach is due to its suitability for our research context which seeks to answer "how" and "why" questions [1]. During the semi-structured interview, the information can be gathered concerning either real-time or retroactive events, since the researcher does not participate in the event but instead only collects information to describe it [1]. Besides, case studies are common to use for exploring the phenomena of study and for discovering relevant factors and characteristics that might apply in similar conditions/situations. Besides, case studies can be used in retroactive events of process change and it is considered as the most appropriate method in this stage of the research evaluation [2]. Another reason, carrying out the case study approach allowing us to know how practitioners do naturally interact with the model and how the model mediate its activities [3, 4]. Also, Darke, et al. [5] stated that the case study approach is suitable to understand the interaction between information technologies related to innovations and organizational context.

Stewart and Shamdasani [6] indicated that there are no general rules as to the optimal number of case studies. They put forward the rationale of working out the number of studies according to the homogeneity of the potential population, and then each of the research application. Furthermore, they suggest that one case study may well be enough. In brief, the widest accepted range seems to fall between 2 to 4 as the minimum and 10, 12, or 15 as the maximum [7]. In this research five real case studies were conducted in the Malaysian SME software development industry to assess the experts' opinion on the development success criteria conformant of the MSME-SPI and to examine whether the MSME-SPI is appropriate for the Malaysian SMEs software development industry. There are five steps to validate the proposed MSME-SPI model are as shown in Figure 2. The steps include, (1) highlight the objectives for model validation, (2) list of the success criteria identified during the initial stages of model development, (3) design a validation instrument to test the success criteria, (4) select an expert panel to reflect the population of experts in SPI implementation; (5) present results of the validation instrument.

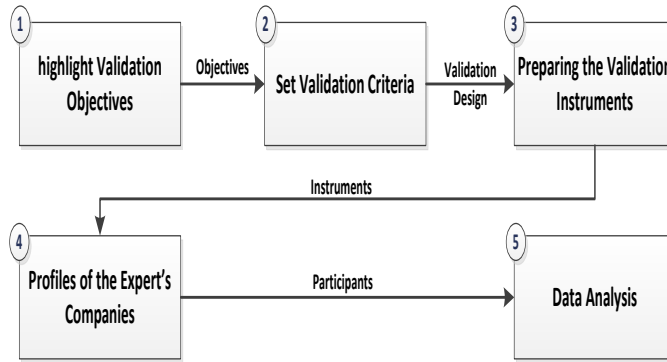

**Fig. 2.** Procedure for validating the MSME-SPI model.

### 2.1 The objective of the validation

As mention earlier, the main objective of this phase is to validate and assess the effectiveness (the success criteria) of the MSME-SPI model in the software development industry. Therefore, the validation was performed by using an expert panel to evaluate criteria include usefulness, verifiability, and structure. An expert in our context is defined as a person who has either several years of experience in implementing any SPI initiatives, or has attained a formal workshop training related to software development and improvement or has responsibilities about software quality within the software development group.

### 2.2 Validation Criteria

The MSME-SPI model is required to verify several issues such as; (1) the usefulness of the model for companies in assessing the strength and weaknesses of their points, (2) the verifiability of the MSME-SPI model practices to be useful for practitioners to implement them, (3) the structure of MSME-SPI model (the suitability of the distribution SFs, BFs and their best practices among different maturity levels) and finally (4) the practices review that is designed under each identified SFs and BFs to cover all practices for any modification or further changes. Therefore, we used these criteria (Usefulness, Verifiability, and Structure,) to evaluate the effectiveness of the MSME-SPI model. These criteria have been gathered, developed, and modified based on existing literature studies used by different researchers [8-12].

*Usefulness,* this criterium emphasizes the satisfaction of Malaysian software practitioners toward the MSME-SPI model. Practitioners should be able to understand the MSME-SPI model clearly without ambiguity to achieve defined objectives according to their expectations [8, 10, 11, 13, 14]. Roles: Perceived usefulness, clear and illuminate the practices, there should be measurable benefits, a model must be appropriated for its objective.

*Verifiability*, this criterium emphasizes the verifiability of the MSME-SPI model in terms of meeting its defined objectives and produces a consistent pattern of results when implemented on different software development environments [8, 10, 11, 13, 14]. Roles: Model must be simple to understand, all terms should be clearly defined, the model must meet its objectives, the model should be provided with a fair assessment.

*Structure*, this criterium emphasizes the key components of the MSME-SPI model to provide clear descriptions and verify the classification of SFs and BFs across its maturity levels [8, 10, 11, 13, 14]. Roles: the structure of the model must be flexible, the structure of the model must be modular, the structure of the model must be transparent, the structure of the model must be suitable with their respective practices, model components should be self-explanatory, model architecture should be unambiguous and functional.

### 2.3 Preparing the Validation Instruments

The main objective of the model validation is to assess the model development success criteria amongst SPI experts in the industry. To achieve the main validation objective, the following information was sought: (1) expert's opinion on the success of the MSME-SPI based on the three success criteria, usefulness, verifiability, and structure. (2) expert's additional comments and suggestions on the MSME-SPI to improve the model. Assessment through case studies, from each company, experts are selected for each case study, who is considered a key person who has either several years of experience in implementing any SPI initiatives. The key expert is communicated face-to-face to explain the model in detail. Assessment through case studies consist of three main stages, which are:

- **Stage one (Training)**: The training stage aims to introduce the MSME-SPI model to the experts. Therefore, short training was held to inform each expert on the basic concepts of SMES-SPI components and its evaluation process.
- **Stage two (Assessment)**: this stage aims to accurately evaluate the company, the experts used MSME-SPI to evaluate their company autonomously without any suggestions or assistance from the researchers. The assessment was performed based on experiences gained from previous projects and assessment results of companies are presented in Appendix K.
- **Stage three (Validation)**: validation stage aims to validate the MSME-SPI model by asking experts to answer based on the three success criteria, usefulness, verifiability, and structure. Therefore, At the end of each case, a feedback session was conducted with the experts to provide feedback about MSME-SPI.

A Consent form was developed for this purpose. The form is divided into three sections. Section one which asks about the demographic information of the participant. This part consists of five questions: Current Position, Working Experience in Malaysian software development industry and Process Improvement (In Years), Experience in the Current Field (e.g., academic or industry), and two close-ended questions related to Knowledge and Experience of the experts towards SPI implementation. Section Two is divided into two parts: Part 1 contains questions asking the participants to evaluate MSME-SPI based on identified criteria as mentioned above. Part 2 contains questions asking the participants to review and comment on critical Success Factors (SFs), critical Barriers Factors (BFs), and their corresponding best practices are. The informal discussion is used during this session to fill out the consent form. The questions in the questionnaires were the close-ended type with a range of response options and open-ended questions ware included to allow respondents to suggest any modification. Each item in Section 2 part 1 used ordinal 5-point Likert scale ("Strongly Agree (SA), Agree (A), Not Sure (SN), Disagree (D), and Strongly Disagree (SD)").

As mentioned earlier, to uncover potential problems in the design of the questionnaire, three software engineering research experts of the Department of Computer & Information Sciences, Universiti Teknologi PETRONAS initially reviewed the case studies documents. The review of the case studies documents was pre-tested to ensure that the documents were formulated. Finally, the necessary amendments suggested by the research experts have been modified. Thus, this helped to improve the completeness and readability of case studies documents. Besides, it guided to refine the validation process.

### 2.4 Profiles of the Expert's Companies

To provide more certainty in this evaluation, five separate case studies were conducted at different Malaysian SME software development environments. These software companies were chosen for case studies because of their familiarity with SPI but could not implement it due to the nature of their software company being an SME. Furthermore, they also gave a special description of their activities in particular which are within our research context. The selected software companies were first identified via Malaysia online directories e.g., www.smecorp.gov.my and www.yellowpages.my. For reasons of privacy, the real names of the respondent companies are not been used, and they are labelled as Company A, B, C, D, and E respectively.

Company A is a local company that is based on software and multimedia engineering. The main business for this company is to provide comprehensive software and multimedia engineering in transforming any software requirements into tangible solutions that meet its costumers' needs and the demand for quality and rapid implementation within affordable cost. Company The main focus is to provide services for its clients in the following areas: Software development production, design, implementation and development, Software development operations and maintenance, Business process consulting and improvement, Software testing consultations, Documentation services for system operations, and Project management Services.

Company B is an experienced software development company specialized in accounting and business intelligence software systems. This company provides consulting services related to services tax and financial reporting standards. Company B is considered as a small software company with less than 50 Employees. The core business of company B is to help all types of businesses and industries company to offer an integrated suite of business management solutions that empower their clients to implement leading-edge business practices, mainly caters to SME sectors. Company B focuses on providing services for industrial companies. For example, Financial services for industries companies, Support companies to comply with Malaysian Goods and Services Tax (GST) and Value-added Tax (VAT) regulations, provide innovative software solutions and technical advisory services, Provide consultation regarding the international financial reporting standards, and Financial Systems integrations.

Company C is a software development outfit with a leading one-stop integrated Information Technology Company listed on the Bursa Malaysia Main Board. Company C is considered as a small software company with less than 50 Employees. The main business domain of Company C is to provide services for banking, e-marketplace, insurance, education, health, government, and finance. The major area of company C is to provide services to meet the need of the customer in the following areas: Design and management of web development projects, E-business services and Solutions, Consultancy and solution integration in IT, Provide software quality for the software industry, and Superior IT solutions for several domains.

Company D is a software development company specialized in developing and managing cloud base business solutions for SME businesses in Malaysia. Company D is considered as a medium software company. The main business domain of Company D is to provide a secure cloud-based business solution including accounting, inventory, customer relationship management, point of sales, and customized cloud solution for businesses to manage core key business operations, with real-time data, easy-to-use views, and role-specific business information. The major area of company D is to provide services to meet the need of the customer in the following areas: accounting software and inventory software, cloud business service provider, and superior cloud business solutions for small and medium businesses.

Company E is a retail solutions software development company in Malaysia. Company E is considered as a small software company with less than 50 Employees. The core business of company E is to is develop and marketing cloud BizSuite such as accounting, CRM, mobile apps, and eCommerce point of sales solutions for retail, food and beverage system, and multi-store management systems. Company E focuses on providing services to meet the need of the customer in the following areas: retail software Solution, Software development operations and maintenance, Business process consulting and improvement, and provide innovative software solutions and technical advisory services.

### 2.5 Data Analysis

A questionnaire survey was used to conduct the data about MSME-SPI validation. Each expert was asked to assess the effectiveness of MSME-SPI against criteria include usefulness, verifiability, and structure. The data analysis process is similar to the process explained and build in the previous data analysis section. The Statistical Package for Social Science (SPSS) version 20 was used for statistical analysis.

## 3      Results and Discussion

A Questionnaire-based survey has been conducted with seven respondents in five real case studies in the Malaysian SME software development industry. Table 1 shows the information regarding the respondents of the case studies.

**Table 1.** Demographic Data of case studies

| ID | Resp.ID | Position | Year of Experience in SD | Year of Experience in SPI |
|---|---|---|---|---|
| A | A-1 | Deployment Manager | 9 | 7 |
| B | B-1 | Manager/ Researcher | 12 | 9 |
| C | C-1 | Project Manager | 11 | 8 |
| D | D-1 | Senior Researcher | 7 | 5 |
|   | D-2 | IT Analyst | 6 | 3 |
| E | E-1 | Head of software development | 9 | 7 |
|   | E-2 | Senior IT officer | 8 | 5 |

To evaluate the ability of the Malaysian SME software development industry to implement SPI in their software activities, the assessment component of MSME-SPI was integrated which was developed based on Motorola Assessment Method [15]. According to Step Five of the assessment tool, to achieve any specific level of MSME-SPI, it is important to cover all critical SFs and BFs that belongs to a particular level and it should have an average score of Seven or higher for it to be considered as 'Strong Implemented' otherwise it will be considered as 'Weak Implemented'.

The respondent of Company A used MSME-SPI and measured the ability of Company A to implement SPI in its software activities. Table 2 reported the summarized results on Company A. The result in Table 6.1 indicated that Company A is at Level 2 of MSME-SPI. Company A has covered all related critical SFs and BFs related to both Level 1 and Level 2 with an average score of each critical SFs and BFs greater than Seven. For Company A to achieve Level 3 in MSME-SPI, it requires addressing three factors which are "Leadership and Employee Involvement", "Staff Learning and Training" and "Lack of Documentation Infrastructure" as their current average score after the assessment for Level 3 in less than Seven. From the results of Company A assessment, more attention should be given to "learning and training", "involvement of staff member", and "documentation infrastructure".

### 3.1      Assessment Results of Company A

To evaluate the ability of the Malaysian SME software development industry to implement SPI in their software activities, the assessment component of MSME-SPI was integrated which was developed based on Motorola Assessment Method [15]. According to Step Five of the assessment tool, to achieve any specific level of MSME-SPI, it is important to cover all critical SFs and BFs that belongs to a particular level and it should have an average score of Seven or higher for it to be considered as 'Strong Implemented' otherwise it will be considered as 'Weak Implemented'.

**Table 2.** Assessment results for company A

| MSME-SPI level | Factors | Score | Result |
|---|---|---|---|
| **Level 0: Immature** | N/A | - | - |
| **Level 1: Realize** | Business Orientation | 8.86 | Strong |
| | Staff Awareness of SPI | 8.80 | Strong |
| | Lack of SPI understanding | 7.20 | Strong |
| **Level 2: Commitment** | Management commitment and support | 8.43 | Strong |
| | Lack of cultural change | 8.00 | Strong |
| **level 3: Manage** | Leadership and employee involvement | 6.27 | Weak |
| | Staff Learning and Training | 4.71 | Weak |
| | Lack of Resources Allocation | 8.50 | Strong |
| | Lack of documentation infrastructure | 5.60 | Weak |
| **level 4: Establish** | Communication support infrastructure | 7.83 | Strong |
| | SPI implementation plan | 7.67 | Strong |
| | Process Tailoring Guidance | 8.60 | Strong |
| | Lack of action knowledge | 4.00 | Weak |
| **Level 5: Optimize** | Review and Feedback | 6.33 | Weak |
| | Lack of Professional Experience and Skills | 8.75 | Strong |

The respondent of Company A used MSME-SPI and measured the ability of Company A to implement SPI in its software activities. Table 2 reported the summarized results of Company A. The result in Table 6.1 indicated that Company A is at Level 2 of MSME-SPI. Company A has covered all related critical SFs and BFs related to both Level 1 and Level 2 with an average score of each critical SFs and BFs greater than Seven. For Company A to achieve Level 3 in MSME-SPI, it requires addressing three factors which are "Leadership and Employee Involvement", "Staff Learning and Training" and "Lack of Documentation Infrastructure" as their current average score after the assessment for Level 3 in less than Seven. From the results of Company A assessment, more attention should be given to "learning and training", "involvement of staff member", and "documentation infrastructure".

### 3.2 Assessment Results of Company B

The respondent of Company B used MSME-SPI to measure the ability of Company B to successfully implement SPI. The assessment results of Company B are reported and summarized in Table 3. The assessment results indicated that Company B is at Level 1 "Realize" of MSME-SPI. While Level 2 was not fully implemented as the average score is less than Seven for "Management Commitment and Support" factor and therefore, Company B must address "Management Commitment and Support" factor to achieve Level 2 "Commitment". Similarly, to achieve Level 3 "Manage" and Level 4 "Establish", Company B need to address four factors, i.e., "Lack of Resources Allocation", "Lack of Documentation Infrastructure", "SPI Implementation Plan", and "Process Tailoring Guidance". The assessment results also showed that Company B usually involves its staff member in its processes and also provides adequate staff development. Furthermore, the assessment result indicated that "Review and Feedback" and "Lack of Professional Experience and Skills" factors are strongly implemented.

**Table 3.** Assessment results for company B

| MSME-SPI level | Factors | Score | Result |
|---|---|---|---|
| **Level 0: Immature** | N/A | - | - |
| **Level 1: Realize** | Business Orientation | 7.10 | Strong |
| | Staff Awareness of SPI | 8.13 | Strong |
| | Lack of SPI understanding | 7.07 | Strong |
| **Level 2: Commitment** | Management commitment and support | 6.95 | Weak |
| | Lack of cultural change | 7.83 | Strong |
| | Leadership and employee involvement | 7.00 | Strong |
| **level 3: Manage** | Staff Learning and Training | 7.67 | Strong |
| | Lack of Resources Allocation | 5.50 | Weak |
| | Lack of documentation infrastructure | 6.20 | Weak |
| | Communication support infrastructure | 7.72 | Strong |
| **level 4: Establish** | SPI implementation plan | 5.83 | Weak |
| | Process Tailoring Guidance | 6.20 | Weak |
| | Lack of action knowledge | 8.20 | Strong |
| **Level 5: Optimize** | Review and Feedback | 7.73 | Strong |
| | Lack of Professional Experience and Skills | 8.42 | Strong |

### 3.3 Assessment Results of Company C

Similarly, to case studies of Company A and B, the respondent of Company C used MSME-SPI and measured the ability of Company C to successfully implement SPI. The assessment results of Company C are reported in Table 4. The assessment results indicated that Company C stands at Level 1 "Realize" of MSME-SPI. The results showed that to achieve Level 2 "Commitment", Company C needs to address all related factors of that level which are "Management Commitment and Support" and "Lack of Cultural Change". To achieve Level 3 "Manage", Level 4 "Establish", and Level 5 "Optimize", Company C must address Six factors, i.e., "Leadership and Employee Involvement", "Staff Learning and Training", "Communication Support Infrastructure", "SPI Implementation Plan", "Review and Feedback", and "Lack of Professional Experience and Skills" factors. The assessment results also indicated that Company C had strongly implemented Four factors in Level 3,4 and 5, i.e., "Lack of Resources Allocation", "Lack of Documentation Infrastructure", "Process Tailoring Guidance" and "Lack of Action Knowledge".

**Table 4.** Assessment results for company C

| MSME-SPI level | Factors | Score | Result |
|---|---|---|---|
| **Level 0: Immature** | N/A | - | - |
| **Level 1: Realize** | Business Orientation | 7.05 | Strong |
| | Staff Awareness of SPI | 7.20 | Strong |
| | Lack of SPI understanding | 7.00 | Strong |
| **Level 2: Commitment** | Management commitment and support | 6.48 | Weak |
| | Lack of cultural change | 6.33 | Weak |
| **level 3:** | Leadership and employee involvement | 6.79 | Weak |
| **Manage** | Staff Learning and Training | 5.86 | Weak |
| | Lack of Resources Allocation | 7.06 | Strong |
| | Lack of documentation infrastructure | 7.73 | Strong |
| **level 4: Establish** | Communication support infrastructure | 6.78 | Weak |
| | SPI implementation plan | 6.67 | Weak |
| | Process Tailoring Guidance | 7.47 | Strong |
| | Lack of action knowledge | 7.73 | Strong |
| **Level 5: Optimize** | Review and Feedback | 6.87 | Weak |
| | Lack of Professional Experience and Skills | 6.67 | Weak |

### 3.4 Assessment Results of Company D

The respondent of Company D used MSME-SPI and measured the ability of their company to successfully implement SPI. The assessment results of Company D are reported in Table 5. The assessment results indicated that Company D stands at Level 1 "Realize" of MSME-SPI. The results showed that to achieve Level 2 "Commitment", Company D need to address "Management Commitment and Support" factors. To achieve Level 3 "Manage", Level 4 "Establish", and Level 5 "Optimize", Company D must address eight factors, i.e., "Leadership and employee involvement", "Staff Learning and Training", "Lack of Resources Allocation", "Lack of documentation infrastructure", "SPI implementation plan", "Process Tailoring Guidance", and "Review and Feedback", "Lack of Professional Experience and Skills". The assessment results also indicated that Company D had strongly implemented Four factors in Level 3,4 and 5, i.e., "Communication support infrastructure" and "Lack of Action Knowledge".

**Table 5.** Assessment results for company D

| MSME-SPI level | Factors | Score | Result |
|---|---|---|---|
| **Level 0: Immature** | N/A | - | - |
| **Level 1: Realize** | Business Orientation | 7.10 | Strong |
| | Staff Awareness of SPI | 7.07 | Strong |
| | Lack of SPI understanding | 7.13 | Strong |
| **Level 2: Commitment** | Management commitment and support | 5.29 | Weak |
| | Lack of cultural change | 7.08 | Strong |
| **level 3: Manage** | Leadership and employee involvement | 5.15 | Weak |
| | Staff Learning and Training | 6.10 | Weak |
| | Lack of Resources Allocation | 5.67 | Weak |
| | Lack of documentation infrastructure | 5.20 | Weak |
| **level 4: Establish** | Communication support infrastructure | 7.56 | Strong |
| | SPI implementation plan | 4.67 | Weak |
| | Process Tailoring Guidance | 5.87 | Weak |
| | Lack of action knowledge | 7.33 | Strong |
| **Level 5: Optimize** | Review and Feedback | 6.13 | Weak |
| | Lack of Professional Experience and Skills | 5.58 | Weak |

### 3.5 Assessment Results of Company E

Similarly, to case studies of the aforementioned companies, the respondent of Company E used MSME-SPI and measured the ability of Company E to successfully implement SPI. The assessment results of Company E are reported in Table 6. The assessment results indicated that Company E was not fully implemented level 1 "Realize" as the average score is less than Seven for "Staff Awareness of SPI" and "Lack of SPI understanding", therefore, company E is required to address these factors ("Staff Awareness of SPI" and "Lack of SPI understanding") to achieve level 1 of MSME-SPI model. Also, the results reported that Level 2 "Commitment" was fully implemented in this company. To achieve Level 3 "Manage", Level 4 "Establish", and Level 5 "Optimize", Company C must address nine factors, i.e., "Leadership and employee involvement", "Staff Learning and Training", "Lack of Resources Allocation", "Lack of documentation infrastructure", "SPI implementation plan", "Process Tailoring Guidance", "Lack of action knowledge", "Review and Feedback", and "Lack of Professional Experience and Skills". The assessment results also indicated that Company E had strongly implemented "Communication support infrastructure" factors in Level 4.

**Table 6.** Assessment results for company E

| MSME-SPI level | Factors | Score | Result |
|---|---|---|---|
| **Level 0: Immature** | N/A | - | - |
| **Level 1: Realize** | Business Orientation | 8.05 | Strong |
| | Staff Awareness of SPI | 6.60 | Weak |
| | Lack of SPI understanding | 6.33 | Weak |
| **Level 2: Commitment** | Management commitment and support | 7.24 | Strong |
| | Lack of cultural change | 7.92 | Strong |
| **level 3: Manage** | Leadership and employee involvement | 6.39 | Weak |
| | Staff Learning and Training | 6.33 | Weak |
| | Lack of Resources Allocation | 5.67 | Weak |
| | Lack of documentation infrastructure | 5.33 | Weak |
| **level 4: Establish** | Communication support infrastructure | 7.06 | Strong |
| | SPI implementation plan | 5.83 | Weak |
| | Process Tailoring Guidance | 5.87 | Weak |
| | Lack of action knowledge | 6.07 | Weak |
| **Level 5: Optimize** | Review and Feedback | 5.60 | Weak |
| | Lack of Professional Experience and Skills | 4.50 | Weak |

### 3.6 Feedback Sessions on the MSME-SPI

After conducting case studies, participants were asked to measure the criteria Usefulness, Verifiability, and Structure of MSME-SPI as explained in the validation criteria Section. In this regard, a questionnaire survey was used to conduct the feedback sessions, to identify any modification or further changes from the respondents of case studies. The Five-point Likert scale is used to identify the effectiveness degree of MSME-SPI "Strongly Agree (SA), Agree (A), Neutral (N), Disagree (D), and Strongly Disagree (SD)".

- *Usefulness Criterion*, the results of key participants of case studies showed that MSME-SPI were completely satisfied concerning the criterion for usefulness. Participants were asked different questions about the usefulness of the model and their feedback were positive. Participants strongly agreed with the results of the assessment. The results of respondents regarding the usefulness criterion are reported. This indicates that all participants completely agreed with the usefulness criterion of different components within the MSME-SPI.
- *Verifiability Criterion*, the participants were asked different questions to rate the verifiability of the model. Also, the participants were asked how well MSME-SPI might be meeting its objectives and how consistent will the model be in the pattern of results when implemented in the different small software development industries. Results of respondents regarding verifiability criterion are reported. which revealed that participants positively agreed with the results of the assessment and their feedback were positive.
- *Structure Criterion*, the main objective of the MSME-SPI structure is to examine the effectiveness of various components of the model and to validate the distribution of critical SFs and BFs across its Six maturity levels. The summarized results of respondents regarding the structure of MSME-SPI are presented that participants positively agreed with the core components of MSME-SPI as a practical robust model for the Malaysian SME software development industry. Furthermore, participants agreed with the maturity level of MSME-SPI to evaluate the ability of the Malaysian SME software development industry to implement SPI.

### 3.7 Modification of the MSME-SPI

Five case studies were conducted to evaluate MSME-SPI in the Malaysian SME software development environment. MSME-SPI was revised based on feedback results from experts in the same field. It was observed that some changes in MSME-SPI structure are important to reflect on case studies results. In this regard, the barrier factor "Lack of culture change" was moved from Level 2 "Commitment" to Level 4 "Establish". Also, one of the respondents recommended moving "Lack of Resources Allocation" factor from Level 3 "Manage" to Level 2 "Commitment". It was also suggested by one of the participants to move "Communication Support Infrastructure" factor from Level 4 "Establish" to Level 3 "Manage". The modified structure of MSME-SPI indicated that Two best practices related to Management Commitment and Support factor are moved, i.e., Practice P2 "The management has committed resources to coordinate SPI activities" moved to Lack of Resources Allocation Factor in Level 2 "Commitment" and Practice P5 "The management is committed to providing training for SPI members" is linked to Staff Learning and Training Factor. Also, one practice related to Awareness of SPI Factor P5 "Awareness procedures is planned to better understand cultural changes" is linked to a Lack of culture change factor. Another expert suggested removing Practices P6 "partners with reliable external expertise provide a mix of resources" from Lack of Resources Allocation Factor. Furthermore, some experts indicated that Practice P5 "SPI implementation plan is constantly reviewed at predefined milestones" should be moved from the SPI implementation plan factor to the Review and Feedback factor at Level 5 "Optimize".

## 4    Conclusion

Validation processes were discussed in detail. A case study was considered as the most appropriate method at this stage of the research evaluation. Therefore, the case study method was used to validate MSME-SPI. In this regard, five separate real-world case studies were conducted at different Malaysian software development environments. Participants' attitudes toward MSME-SPI were positively performed. However, authors recognize that there may be bias in how the proposed model (MSME-SPI) was developed. Nevertheless, the validation serves as a guide for further development and future work to improve the MSME-SPI to assess the Malaysian SMEs software development industry to implement SPI in their software activities.